\documentstyle[12pt,epsfig,amssymb]{article}
\begin{document}
\setlength{\baselineskip}{0.75cm}
\setlength{\parskip}{0.45cm}
\begin{titlepage}
\begin{flushright}
ANL-HEP-PR-96-15 \\
RAL-TR-96-057 \\
July 1996
\end{flushright}
\vspace*{1.5cm}
\begin{center}
\Large{
{\bf Inclusive Prompt Photon Production in Polarized $pp$ Collisions at 
HERA-$\vec{{\rm N}}$}}
\vskip 2.5cm
{\large L.E. Gordon}  \\
\vspace*{0.6cm}
\normalsize
High Energy Physics Division, Argonne National Laboratory, \\
Argonne IL 60439, USA \\
\vspace*{1cm}
{\large W. Vogelsang}  \\
\vspace*{0.6cm}
\normalsize
Rutherford Appleton Laboratory, \\ 
Chilton DIDCOT, Oxon OX11 0QX, England
\end{center}
\vskip 2cm
\begin{center}
{\bf Abstract} \\
\end{center}
We present a NLO study of inclusive polarized prompt photon production
in a conceivable fixed target $pp$ mode of HERA with longitudinally 
polarized protons at $\sqrt{s}=39$ GeV. We analyze the sensitivity of 
the corresponding double spin asymmetry to the proton's polarized gluon 
distribution $\Delta g$ and estimate the expected statistical precision 
in its determination. The main theoretical uncertainties in the predictions 
are examined.
\vskip 2cm
\end{titlepage}
It has recently become possible to perform a complete and consistent study 
of longitudinally polarized deep-inelastic scattering (DIS) in 
next-to-leading order (NLO) of QCD, since the spin-dependent two-loop 
splitting functions, needed for the NLO evolution of the polarized parton 
distributions, have become available \cite{MVN,wv}. A first such 
phenomenological NLO analysis, taking into account all available experimental 
data on polarized DIS \cite{voss} has been presented in~\cite{grsv}, followed
by the analyses~\cite{bfr,gs}. The studies of \cite{grsv,gs} have shown 
that present polarized DIS data are still quite far from providing accurate 
knowledge about the nucleon's spin-dependent sea quark and gluon distributions. 
This holds true, in particular, for the polarized gluon density $\Delta g$, 
the $x$-shape of which seems to be hardly constrained at all \cite{grsv,gs}
by the DIS data, even though a tendency towards a sizeable positive 
{\em total} gluon polarization, $\int_0^1 \Delta g(x,Q^2=4 \; \mbox{GeV}^2) 
dx \gtrsim 1$, was found \cite{grsv,bfr,gs}. Thus, there is clearly some 
need for independent information on $\Delta g$. For this purpose, it seems 
expedient to look at processes for which $\Delta g$ enters in leading 
order (LO) already, rather than as a NLO correction as for the 
spin-dependent DIS structure function $g_1$. One of such processes 
is inclusive large-$p_T$ prompt photon production in collisions of 
longitudinally polarized protons, $\vec{p}\vec{p} \rightarrow \gamma X$ 
\cite{berger,cont,wir,wirphen}. In the unpolarized 
case where this process has been studied in a huge number of experiments 
it has been an invaluable tool for pinning down the proton's unpolarized 
gluon distribution $g(x,Q^2)$ \cite{aur,mrsa,ctq,vvdg}. Hence prompt photon 
production with polarized beams seems a promising source for obtaining 
information on $\Delta g$. 

It is being discussed as one future option for HERA to polarize its 
$820$ GeV proton beam. If this can be achieved, one could use the 
beam in a fixed target experiment, scattering it off an internal 
polarized nucleon target. This conceivable constellation, 
dubbed 'Phase II' of HERA-$\vec{{\rm N}}$ \cite{nowak}, would yield 
$\sqrt{s} \approx 39$ GeV and thus could provide information 
complementary to that obtained from planned similar spin physics experiments 
at much higher energies at the RHIC collider \cite{rhic}. Theoretical 
predictions for polarized prompt photon production at $\sqrt{s}\approx 
40$ GeV have been made in the past \cite{cont,wirphen}, 
taking into account the spin-dependent 'direct' subprocess cross sections 
for $ab\rightarrow \gamma X$ ($a,b=q,\bar{q},g$) including their full
NLO QCD corrections as calculated in \cite{cont,wir}. 
From the experience in the unpolarized case, the inclusion of 
NLO corrections is expected to be quite important in order to make reliable 
predictions. The main shortcoming of the studies \cite{cont,wirphen} was, 
however,
that spin-dependent parton densities evolved only in LO had to be used at 
that time. Having sets of NLO polarized parton distributions available now, we 
can obviously put the corresponding predictions as well as the assessment 
of their theoretical uncertainties such as their scale dependence 
on a much firmer basis, which is the purpose of this paper. The only remaining 
drawback here is that the fragmentation contribution to 
the polarized prompt photon cross section still cannot be calculated 
in NLO since the NLO corrections to the underlying polarized subprocesses 
$ab\rightarrow c X$ ($a,b,c=q,\bar{q},g$) are not yet known. On the other 
hand, in the unpolarized case the fragmentation piece is known
to be subdominant -- even though not negligible -- at fixed target energies
(see, e.g., \cite{vvdg}). The fragmentation contribution to the 
polarized cross section which was omitted altogether 
in \cite{cont,wirphen}, will be included on a LO basis in 
this paper. Even though this is not strictly consistent in the framework of a
NLO calculation, it is the best 'state-of-the-art' procedure and also appears 
reasonable in view of the fact that the $K$-factor $K^{frag} \equiv 
\sigma_{NLO}^{frag}/\sigma_{LO}^{frag}$ for the fully inclusive fragmentation 
part generally turns out to be very close to unity when calculated for the 
unpolarized case.

The analyses of \cite{grsv,gs} provide several different sets of LO and NLO 
spin-dependent parton densities, all of which are in very good agreement 
with the existing polarized DIS data but differ mainly in the 
$x$-shapes of their polarized gluon distributions. We are therefore in the 
position to study the sensitivity of polarized prompt photon production 
to $\Delta g (x,Q^2)$ on the basis of consistent NLO 
sets of parton distributions 
that include all experimental information presently available from DIS, but 
also reflect the full freedom concerning $\Delta g$ left by those data. 
The latter is illustrated in Fig.~1 which compares the gluon distributions 
of various NLO sets of \cite{grsv,gs} in the $x$-range dominantly probed 
by prompt photon production at $\sqrt{s}=39$ GeV and $p_T\geq3$ GeV.
The sets we will use in our study 
are the NLO 'valence' set of the 'radiative parton model analysis' 
\cite{grsv}, which corresponds to the best-fit result of that paper
(hereafter referred to as 'fitted $\Delta g$' scenario), and
two other sets of \cite{grsv} which are based on either assuming 
$\Delta g (x,\mu^2) = g(x,\mu^2)$ or $\Delta g(x,\mu^2)=0$ at the low input 
scale $\mu$ of \cite{grsv}, where $g(x,\mu^2)$ is the unpolarized NLO GRV 
\cite{grv} input gluon distribution. These two sets will be called
'$\Delta g=g$ input' and '$\Delta g=0$ input' scenarios, respectively, 
in what follows. It turns out that in the $x$-range explored here 
the NLO sets A,B of \cite{gs} have gluon distributions quite 
similar to those of the '$\Delta g=g$ input' and 'fitted $\Delta g$' 
scenarios of \cite{grsv}, respectively. Only the gluon of set C of 
\cite{gs} ('GS C') is qualitatively different since it has a 
substantial negative polarization 
at large $x$. We will therefore also use this set in our calculations.
A graph similar to Fig.~1 could be shown for the corresponding LO gluon 
distributions of \cite{grsv,gs} (see, e.g., \cite{sv}). The fact that 
\cite{grsv,gs} have provided both consistent LO {\em and} NLO parton sets in 
all cases, enables us to more reliably study the perturbative stability of 
the cross sections and asymmetries. This is again a major improvement 
with respect to the previous studies \cite{cont,wirphen}.

The quantity to be studied in prompt photon experiments with polarized
beam and target is the double-spin asymmetry 
\begin{equation}
A_{LL} \equiv \frac{d\sigma^{++} - d\sigma^{+-}}
{d\sigma^{++} + d\sigma^{+-}}
\equiv \frac{d\Delta \sigma}{d\sigma} \; ,
\end{equation}
where $d\sigma^{++}$ ($d\sigma^{+-}$) denotes the cross section for the 
prompt photon being produced by protons with same (opposite) helicities.
Two types of processes contribute to the prompt photon production cross
section: the so-called `direct' piece, where the photon is emitted via
a pointlike (direct) coupling to a quark, and the fragmentation piece,
in which the photon originates from the fragmentation of a final state
parton. The cross section for the fully inclusive production of a prompt 
photon with momentum $p_{\gamma}$ thus schematically reads
\begin{eqnarray}  \label{wq}
d\Delta \sigma  &\equiv& \frac{1}{2} \left( d\sigma^{++} - d\sigma^{+-} 
\right) \equiv d\Delta \sigma_{dir}+d\Delta \sigma_{frag} \\
&=&\sum_{a,b=q,\bar{q},g}\int dx_a dx_b
\Delta f_a(x_a,\mu_F^2) \Delta f_b(x_b,\mu_F^2) 
\Bigg[ d\Delta \sigma_{ab}^{\gamma}(p_{\gamma}, x_a,x_b, \mu_R,
\mu_F,M_F) \nonumber \\
&&+ \left. \sum_{c=q,\bar{q},g} \int \frac{dz}{z^2}
d\Delta \sigma_{ab}^c (p_{\gamma},x_a,x_b,z,\mu_R,\mu_F,M_F)
D^{\gamma}_c (z,M_F^2)\right] \; , \nonumber
\end{eqnarray}
where the $d\Delta \sigma_{ab}^i$ represent the spin-dependent 
subprocess cross sections for partons $a,b$ producing a particle $i$ 
($i=\gamma$, $q$, $\bar{q}$, $g$), integrated over the full phase space 
of all other final state particles. The polarized parton distributions are 
defined as 
\begin{equation}  \label{fdef}
\Delta f_i(x,\mu_F^2) = f_i^+ (x,\mu_F^2) - f_i^- (x,\mu_F^2) \; ,
\end{equation} 
with $f_i^+ (x,\mu_F^2)$ ($f_i^- (x,\mu_F^2)$) denoting the number 
density of parton-type $i$ with momentum fraction $x$ and positive (negative) 
helicity in a proton with positive helicity at scale $\mu_F$. Furthermore, 
in (\ref{wq}) $D_c^{\gamma}(z,M_F^2)$ is the (unpolarized) photon 
fragmentation function at scale $M_F$, $z$ being the fraction of energy of 
the fragmenting parton $c$ transferred to the photon. We note that even 
in the polarized case the photon fragmentation functions are always the 
unpolarized ones since the polarization of the outgoing photon is not 
observed. Despite the fact that its corresponding partonic 
subprocesses are of order $\alpha_s^2$, the fragmentation contribution is 
present already in LO since the parton-to-photon fragmentation functions are
effectively of order $\alpha_{e.m.}/\alpha_s$ in perturbative QCD. 
As made explicit in (\ref{wq}), the cross section in any fixed order of 
perturbation theory depends on unphysical scales which have to be introduced 
in the procedure of renormalization ($\mu_R$) and of factorization of initial 
($\mu_F$) and final ($M_F$) state mass singularities. Unless stated 
otherwise, we will choose $\mu_R=\mu_F=M_F=p_T/2$ in what follows.

The corresponding expressions for the unpolarized cross section, which we 
need to calculate the spin asymmetry $A_{LL}$, can be obtained from 
(\ref{wq}),(\ref{fdef}) by omitting the '$\Delta$' and taking sums 
instead of the differences on the rhs. In this case, all cross sections 
$d \sigma_{ab}^i$ ($i=\gamma$, $q$, $\bar{q}$, $g$) are known to NLO accuracy, 
the corrections to the direct and fragmentation subprocess cross sections
having been calculated in the $\overline{\mbox{MS}}$ scheme in 
\cite{boo,wir} and \cite{acgg}, respectively. For consistency, 
when including these NLO corrections, we have to use NLO unpolarized parton 
densities and photon fragmentation functions. For the latter we use 
those of GRV \cite{grvfrag} throughout. The specific choice for the
protonic parton densities turns out to be rather immaterial, all 
modern parametrizations leading to essentially the same results.
The sets of spin-dependent NLO parton densities we want to use have in 
each case been set up in relation to some underlying 'reference' 
set of NLO {\em un}polarized densities: For the distributions of 
\cite{grsv} this has
been the GRV set \cite{grv}, whereas the MRS(A') densities \cite{mrsa} 
have been adopted in \cite{gs}. For definiteness, we will choose the GRV NLO 
parton distributions \cite{grv} as our 'standard' for the unpolarized case, 
but use the MRS(A') set \cite{mrsa} when employing the polarized  NLO
densities of \cite{gs}. This also implies using the values for 
$\Lambda_{\overline{\rm{MS}}}^{(f)}$ ($f$ being the number of 
flavors) as implemented in the respective NLO sets, e.g.
$\Lambda_{\overline{\rm{MS}}}^{(f=4)}=200$ MeV for the GRV and 
$\Lambda_{\overline{\rm{MS}}}^{(f=4)}=231$ MeV for the MRS(A') set, in
the NLO expression for the strong coupling $\alpha_s$:
\begin{equation}
\label{as}
\frac{\alpha_s(Q^2)}{4\pi} \simeq \frac{1}{\beta_0 \ln Q^2/
\Lambda_{\overline{\rm{MS}}}^2} - \frac{\beta_1}{\beta_0^3}
\frac{\ln \ln Q^2/\Lambda_{\overline{\rm{MS}}}^2}
{\left(\ln Q^2/\Lambda_{\overline{\rm{MS}}}^2\right)^2} \; ,
\end{equation}
where $\beta_0=11-2 f/3$, $\beta_1=102-38 f/3$. As provided for in the
sets of \cite{grv,mrsa}, the number of flavors increases when crossing a 
heavy flavor threshold, and the value for $\Lambda_{\overline{\rm{MS}}}^{(f)}$ 
changes as a result of the continuity of $\alpha_s$ across the threshold.
We follow this prescription also for the explicit $f$ appearing in the 
NLO subprocess cross sections. We neglect, however, the genuine charm 
(and, of course, bottom) contributions to the polarized and unpolarized  
cross sections which are tiny at $\sqrt{s}=39$ GeV \cite{vvdg}.

In the polarized case, next-to-leading order QCD corrections have
been calculated for the spin-dependent direct subprocess cross sections
$d\Delta \sigma_{ab}^{\gamma}$ \cite{cont,wir}. 
We emphasize that care has to be taken when combining the NLO sets of 
polarized parton distributions of \cite{grsv,gs} with the NLO expressions for 
the $d\Delta \sigma_{ab}^{\gamma}$ of \cite{cont,wir} to avoid a mismatch 
in the factorization schemes used. The two-loop splitting functions of 
\cite{MVN,wv}, which were employed in \cite{grsv,gs}, have been calculated in 
dimensional regularization in the conventional $\overline{\mbox{MS}}$ scheme.
To be more precise, use of dimensional regularization in such a calculation
implies to choose a prescription for dealing with the Dirac matrix 
$\gamma_5$ and the Levi-Civita tensor $\epsilon_{\mu\nu\rho\sigma}$ which 
enter as projectors onto definite helicity states of the involved particles.
In \cite{MVN} the 'reading point' method of \cite{korn} with a fully 
anticommuting $\gamma_5$ was chosen, whereas \cite{wv} adopted the original 
definition for $\gamma_5$ of \cite{hvbm} (HVBM scheme) which is widely 
considered to be the most consistent prescription. It turned out that both 
calculations \cite{MVN,wv} arrived at the same final result for the polarized 
two-loop splitting functions. The polarized NLO parton distributions of 
\cite{grsv,gs} therefore refer to the conventional $\overline{\mbox{MS}}$ 
factorization scheme {\em in combination} with the HVBM prescription 
\cite{hvbm} (or the one of \cite{korn}) for $\gamma_5$, and the NLO 
$d\Delta \sigma_{ab}^{\gamma}$ have to be known in the {\em same} 
scheme to make it sensible to use them in conjunction with the NLO partons of
\cite{grsv,gs}. In fact, the calculation of \cite{wir} of the NLO 
corrections to the $d\Delta \sigma_{ab}^{\gamma}$ has been performed 
using the HVBM prescription, which makes the results suitable for our 
purposes. We note, however, that in \cite{wir} a slight deviation from the
$\overline{\mbox{MS}}$ scheme was made by factorizing certain finite 
'collinear' terms, arising from the non-fourdimensional parts of the polarized 
LO splitting functions as calculated in the HVBM scheme, into the NLO 
spin-dependent parton densities along with the collinear singularities 
('$\overline{\mbox{MS}}_P$ scheme'). It is straightforward to invert 
this procedure, i.e., to bring back the results of \cite{wir} to the
conventional $\overline{\mbox{MS}}$ scheme. Details of this are given in
the appendix. We note that we refrain from using the results of \cite{cont}
for the NLO corrections to the $d\Delta \sigma_{ab}^{\gamma}$ since
it is not obvious whether the $\gamma_5$ scheme used in \cite{cont} provides 
a consistent regularization and can be related in any way to the HVBM 
prescription.

Unfortunately, the NLO corrections to the polarized fragmentation subprocess
cross sections $d\Delta \sigma_{ab}^i$ ($i=q$, $\bar{q}$, $g$)
are still unknown as mentioned above. For this reason we stick to a pure LO 
calculation for this contribution, using the corresponding spin-dependent 
LO subprocess cross sections of \cite{bab} along with the respective LO sets of 
polarized parton distributions of \cite{grsv,gs}, and the LO expression 
for $\alpha_s$ (as entailed in (\ref{as}) by dropping the $\beta_1$ term)
with \cite{grv} $\Lambda_{LO}^{(f=4)}=200$ MeV. For the photon 
fragmentation functions we use the LO set of \cite{grvfrag} in this case.

Fig.~2a shows the NLO predictions for the spin-dependent cross 
sections $d\Delta \sigma/dp_T d \eta$ as functions of the prompt
photon's transverse momentum $p_T$ at c.m.s. rapidity $\eta=0$
for the four different sets of polarized parton distributions. We also
display the unpolarized NLO cross section. In Fig.~2b we show the 
asymmetries $A_{LL}$ corresponding to Fig.~2a. As becomes obvious, 
$A_{LL}$ depends strongly on the size and shape of $\Delta g$ even at 
large $p_T$. When artificially setting $\Delta g(x,\mu_F^2) \equiv 0$
one finds that the asymmetry becomes very small and negative for all $p_T$
and all four sets of polarized parton distributions. Thus the differences 
between the results in Fig.~2b are indeed due to the polarized gluon 
distribution employed. The small negative 'offset' in the asymmetry obtained 
for $\Delta g(x,\mu_F^2) \equiv 0$ turns out to be be due to the LO 
annihilation process $q\bar{q} \rightarrow \gamma g$ and NLO corrections 
involving only incoming quarks and is mainly responsible   
for the fact that the asymmetries for the '$\Delta g=0$ input' and the 
'fitted $\Delta g$' scenarios of \cite{grsv} are much closer to each other 
at large $p_T$ than those for the 'fitted $\Delta g$' and the 
'$\Delta g=g$ input' scenarios. We also note that the full (LO)
fragmentation contribution to the polarized cross section is positive
for all parton sets (apart from set C of \cite{gs} at small to medium $p_T$)
and not very sensitive to $\Delta g$ for $p_T \gtrsim 7$ GeV. 

Figs.~2b,d show the same quantities as Figs.~2a,c, but now as functions 
of $\eta$ at $p_T=6$ GeV. We note that the rapidity range $-1.5 \leq \eta
\leq 1.5$ shown is equivalent to laboratory angles of $0.9^{\circ} \leq
\Theta_{lab} \leq 17^{\circ}$, which roughly corresponds to the 
range expected to be 
accessible in the HERA-$\vec{{\rm N}}$ experiment at this $p_T$ 
\cite{nowak}. Again, the asymmetries in Fig.~2d show strong sensitivity 
to $\Delta g$, even becoming slightly better at large $|\eta|$. 

We have included in the asymmetry plots in Figs.~2b,d the expected statistical
errors $\delta A_{LL}$ at HERA-$\vec{{\rm N}}$ which can be estimated from 
\begin{equation}  \label{aerr}
\delta A_{LL} = 0.17/\sqrt{\sigma \; ( \mbox{pb} )}  \; .
\end{equation}
This relation has been determined in \cite{nowak} assuming an integrated 
luminosity of 240 pb$^{-1}$ and beam and target polarizations $P_B=0.6$,
$P_T=0.8$. It includes an overall trigger and reconstruction efficiency 
of $50\%$ but no acceptance correction. The error
bars in Figs.~2b,d have been obtained using our unpolarized NLO $d\sigma/
dp_T d\eta$ in (\ref{aerr}), integrated over bins of $\Delta \eta=1$,
$\Delta p_T=1$ GeV. The bars have been plotted at the weighted centers
of the bins. It becomes obvious that the asymmetry should be measurable 
by HERA-$\vec{{\rm N}}$ for $p_T \leq 7$ GeV and for almost all accessible 
$\eta$, and one should be able to distinguish between different scenarios
for $\Delta g (x,\mu_F^2)$ at $0.1 \leq x \leq 0.4$.

The remainder of this paper is devoted to an examination of the reliability
of the predictions of these very positive findings. We will first discuss the 
actual size of the NLO effects that we have included, and then try to assess 
the main uncertainties in the predictions.

Figs.~3a,c show the $K$-factors 
\begin{equation} \label{kfac}
K \equiv \frac{d(\Delta)\sigma_{NLO}}{d(\Delta)\sigma_{LO}}
\end{equation}
for the unpolarized and the polarized cross sections for our three sets 
of spin-dependent parton distributions of \cite{grsv}. From now on, we do 
not show the corresponding results for set C of \cite{gs} to avoid a 
proliferation of curves. For the $K$-factors the LO cross sections 
in (\ref{kfac}) have been calculated consistently, i.e. with LO parton
distributions and photon fragmentation functions and LO $\alpha_s$.
The $K$-factors turn out to be quite close to unity in the experimentally 
accessible ranges of $p_T$ and $\eta$. Only for the '$\Delta g=0$ input' 
scenario is the $K$-factor much less than unity around $\eta=0$ for all $p_T$,
indicating rather significant NLO corrections in this case. Figs.~3b,d
compare the asymmetries $A_{LL}$ in NLO (as already shown in Figs.~2b,d) 
and LO. By comparison of Figs.~3a,c and 3b,d it can be seen that the 
asymmetries are to some extent less influenced by the NLO corrections than 
the individual polarized and unpolarized cross sections. This is again 
not true for the '$\Delta g=0$ input' scenario in Fig.~3b, whose asymmetry
changes quite a lot for all $p_T$ when going from LO to NLO.
This feature for a scenario with a small $\Delta g$ was already observed
in \cite{cont,wirphen} and is due to sizeable negative contributions from 
direct genuine NLO processes like $qq' \rightarrow \gamma qq'$. 

One major uncertainty in our predictions is expected to come from the
fragmentation contribution since the parton-to-photon fragmentation
functions are experimentally unmeasured so far, even though very sensible
theoretical predictions for them are available \cite{grvfrag,ACFGP}
(see also \cite{vvdg}). In our case, the uncertainty is even larger
since, as discussed earlier, we have to stick to a pure LO calculation
for the fragmentation contribution in the polarized case, rather than 
including it on a NLO basis as would be required by consistency.
It seems likely, however, that the individual $K$-factor for fully inclusive
polarized fragmentation is close to unity, just as it turns out to be in the 
unpolarized case\footnote{We note that this is no longer the case if the 
fragmentation contribution to the {\em isolated} prompt photon cross 
section, as measured at very high-energy colliders, is considered. 
Here the unpolarized fragmentation piece turns out to be substantially 
larger at NLO than when calculated in LO \cite{gvisol,ggrv}. This 
feature could set a limitation to the accuracy of similar theoretical 
predictions for polarized prompt photon production at RHIC where the 
introduction of some experimental isolation criterion will almost certainly
be necessary.}. Further investigation of this 
issue is needed in future, requiring a calculation of the NLO QCD corrections 
to the spin-dependent parton-parton scattering cross sections. To assess the 
importance of fragmentation, we show in Figs.~4a,c the ratios
\begin{equation} \label{ratio}
R \equiv \frac{d(\Delta)\sigma_{frag}}{d(\Delta) \sigma_{dir}+
d(\Delta) \sigma_{frag}}
\end{equation}
in NLO (apart, of course, from the polarized $d\Delta \sigma_{frag}$ which is 
LO). It turns out that the fragmentation contribution is generally of 
${\cal O}(20 \%)$ at $\sqrt{s}=39$ GeV and in the ranges of $p_T$, $\eta$ 
shown 
(see also \cite{vvdg} for the unpolarized case), but that it is relatively
much larger for the '$\Delta g=0$ input' scenario around $\eta=0$, where its 
importance even rises with rising $p_T$. This surprising result is, however, in 
accord with our previous finding that the fragmentation part is positive and 
rather independent of $\Delta g$ at large $p_T$. Thus in the '$\Delta g=0$ 
input' scenario, where the direct contribution to the cross section is 
mainly due to the small and negative $q\bar{q} \rightarrow \gamma g$ 
annihilation process, fragmentation plays an important role even at large 
$p_T$ since cancellations between the direct and the fragmentation pieces 
occur. Even though it does not seem likely that the finding of general 
smallness of the asymmetry $A_{LL}$ for this scenario would become invalid
if fragmentation could be included at NLO, it means 
that our predictions for the '$\Delta g=0$ input' scenario are the least 
certain ones since a strong deviation of the individual $K$-factor
for polarized fragmentation from unity would affect the predictions for this 
scenario most.

On the other hand, we find a very reassuring result in this context when 
examining another major source of uncertainty, namely the dependence 
of the results on the unphysical scales $\mu_R$, $\mu_F$ and $M_F$. 
It turns out that the dependence of {\em all} polarized cross sections
on the fragmentation scale $M_F$ alone is already {\em extremely} weak, 
even though fragmentation is only included in LO. This feature, which was 
also found in \cite{vvdg} for the fully inclusive unpolarized cross 
section (where NLO fragmentation was used) might be evidence for indeed very 
mild NLO corrections to polarized fragmentation. 

In contrast to this, the dependence of the polarized and unpolarized NLO cross 
sections on the renormalization and initial-state factorization scales is 
very strong. For instance, changing the scales to $\mu_R=\mu_F=M_F=2 p_T$ 
(we include the change of $M_F$ here even though it has hardly any influence), 
the unpolarized NLO cross section decreases by $\gtrsim 50\%$ almost uniformly 
over the whole ranges of $p_T$ and $\eta$ (see \cite{vvdg} for a closer 
examination of the notorious scale dependence of the unpolarized inclusive 
prompt photon cross section). Nevertheless, the asymmetries $A_{LL}$, which 
will be the quantities actually measured, remain quite untouched by scale
changes, as can be seen from Figs.~4b,d. This finding is very important 
since it warrants the genuine sensitivity of $A_{LL}$ to $\Delta g$, 
implying that despite the sizeable scale dependence of the cross sections it 
still seems a reasonable and safe procedure to compare theoretical predictions 
for the asymmetry with future data and to extract $\Delta g$ from such 
comparisons.

In conclusion, we have presented a careful NLO analysis of 
HERA-$\vec{{\rm N}}$'s capability to measure the nucleon's polarized gluon 
distribution $\Delta g$ in inclusive prompt photon production with polarized 
beam and target. The corresponding double spin asymmetry $A_{LL}$ shows 
strong sensitivity to $\Delta g$, and its measurement as well as a 
distinction between various possible scenarios for $\Delta g(x,\mu_F^2)$ in 
the range $0.1 \leq x \leq 0.4$ should be possible experimentally. 
We have also assessed the uncertainties in the theoretical predictions 
for $A_{LL}$ which appear to be under control.
\section*{Acknowledgements}
We are thankful to W.-D. Nowak for helpful comments. This work  
was supported in part by the US Department of Energy, Division of
High Energy Physics, Contract number W-31-109-ENG-38. 
\section*{Appendix}
In this appendix we list the changes in the NLO corrections to the 
direct part of polarized prompt photon production which are to be made to 
transform the results of \cite{wir} from the '$\overline{\mbox{MS}}_P$ scheme' 
back to the conventional $\overline{\mbox{MS}}$ scheme. They are entirely
due to terms $\sim \epsilon (1-x)$ in the polarized $n=4-2\epsilon$ 
dimensional LO splitting functions $\Delta P_{ij}^{(n)}(x)$ ($i,j=q,g$)
as calculated in the HVBM scheme, which were absorbed into the spin-dependent 
parton densities in \cite{wir}. We emphasize that we must {\em not} undo
this procedure of \cite{wir} for the case of $\Delta P_{qq}^{(n)}(x)$ 
since in this case the subtraction of the adjacent terms $\sim \epsilon (1-x)$ 
has also been performed in the calculation of the spin-dependent NLO 
splitting functions where it is demanded by the conservation of the
axial non-singlet current \cite{MVN,wv}. We therefore only have to
reintroduce the effects of the terms $\sim \epsilon (1-x)$ in 
$\Delta P_{ij}^{(n)}(x)$ ($\{ij\} \neq \{qq\}$) (as listed in \cite{wir}) 
into the polarized NLO cross sections. As a consequence of this, only 
the NLO cross sections for some subprocesses need to be 
changed. Furthermore only the coefficients $\Delta c_{13}$ of \cite{wir}, 
corresponding to the terms involving no logarithms and no distributions, 
are affected. The following terms have to be {\em added} to the coefficients 
$\Delta c_{13}$ in \cite{wir} to transform the corresponding polarized NLO 
subprocess cross sections back to the conventional $\overline{\mbox{MS}}$ 
scheme:    
\begin{description}
\item[{\bf $qg\rightarrow \gamma qg$} :]
$$ - \frac{v^2 (1 - w)}{X^2} \left( 2 (1 + X) v^2 w^2 + 
     \frac{C_F}{N_C} ( 1 - 2Xvw) \right)$$
\item[{\bf $gg\rightarrow \gamma q\bar{q}$} :]
$$ \frac{1}{N_C} (2 + Y) v^2  (1 - w) $$
\item[{\bf $qq\rightarrow \gamma qq$} :]
$$ \frac{C_F}{N_C X^2}  (1 - w) \left( (1 - 2X) (1 + v) v_1^2
   - v^2 (X v^2 + v^3 + v_1) w^2 \right)$$
\item[{\bf $q\bar{q}\rightarrow \gamma q\bar{q}$} :]
$$ \frac{C_F}{N_C X^2}  (1 - w) \left( (1 - 2X) (1 + v) v_1^2
   - v^2 (X v^2 + v^3 + v_1) w^2 \right)$$
\item[{\bf $qq'\rightarrow \gamma qq'$} :]
$$ -e_q^2 \frac{C_F}{N_C X^2} (1 + X) v^4 (1-w) w^2 
-e_q^{'2} \frac{C_F}{N_C} (1 + v) v_1^2 (1-w) \; \; ,$$
\end{description}
where $C_F=4/3$, $N_C=3$ and the variables $v$, $v_1$, $w$, $X$, $Y$ are as
defined in \cite{wir}. For the last process, $e_q$ and $e_q'$ denote the
charges of $q$ and $q'$, respectively.
The coefficients $\Delta c_{13}$ for the processes $q\bar{q}\rightarrow 
\gamma gg$ and $q\bar{q}\rightarrow \gamma q'\bar{q}'$ remain unchanged.

\newpage
\section*{Figure Captions}
\begin{description}
\item[Fig.1] Gluon distributions at $Q^2=10$ GeV$^2$ of the four NLO
sets of polarized parton distributions used in this paper. The dotted
line refers to set C of \cite{gs}, whereas the other distributions
are taken from \cite{grsv} as described in the text. For comparison
we also show the unpolarized NLO gluon distribution of \cite{grv}.
\item[Fig.2] {\bf a:} $p_T$-dependence of the NLO polarized cross section 
for inclusive prompt photon production at HERA-$\vec{{\rm N}}$ for the four 
sets of spin-dependent parton distributions. The results are presented at 
c.m.s. rapidity $\eta=0$. The cross section for set C
of \cite{gs} has been multiplied by $-1$. For comparison we also
show the unpolarized NLO cross section. The scales have been chosen to be
$\mu_R=\mu_F=M_F=p_T/2$.
{\bf b:} Asymmetries $A_{LL}$ corresponding to {\bf a} (here the result
for set 'GS C' \cite{gs} is shown with its actual sign). The expected 
statistical errors indicated by the bars have been calculated according
to Eq.~(\ref{aerr}) and as explained in the text.
{\bf c,d:} Same as {\bf a,b}, but for the $\eta$-dependence at 
$p_T=6$ GeV.
\item[Fig.3] {\bf a:} $p_T$-dependence of the $K$-factors (as defined in
Eq.~(\ref{kfac})) for the unpolarized and the polarized cross sections at
$\eta=0$. The polarized results are shown for the three sets of parton 
distributions taken from \cite{grsv} (see text); line drawings are as in 
Fig.~2. {\bf b:} Comparison of the LO and NLO asymmetries
$A_{LL}$ for the three sets of polarized parton distributions taken 
from \cite{grsv} (see text). The NLO curves are as already shown in Fig.~2b.
{\bf c,d:} Same as {\bf a,b}, but for the $\eta$-dependence at  
$p_T=6$ GeV.
\item[Fig.4] {\bf a:} $p_T$-dependence of the ratios $R$ (as defined in
Eq.~(\ref{ratio})) for the unpolarized and polarized cross sections at
$\eta=0$. The polarized results are shown for the three sets of parton
distributions taken from \cite{grsv} (see text); line drawings are as in 
Fig.~2. {\bf b:} The scale dependence of the NLO asymmetries 
$A_{LL}$ for the three sets of polarized parton distributions
taken from \cite{grsv} (see text). The curves for the scales 
$\mu_R=\mu_F=M_F=p_T/2$ are as already shown in Fig.~2b and are 
displayed in the same line drawings. For each set the dotted (long-dashed) 
line corresponds to the scales $p_T$ ($2p_T$).
{\bf c,d:} Same as {\bf a,b}, but for the $\eta$-dependence at
$p_T=6$ GeV.  
\end{description}
\newpage 
\pagestyle{empty}

\vspace*{0.0cm}
\hspace*{-1.4cm}
\epsfig{file=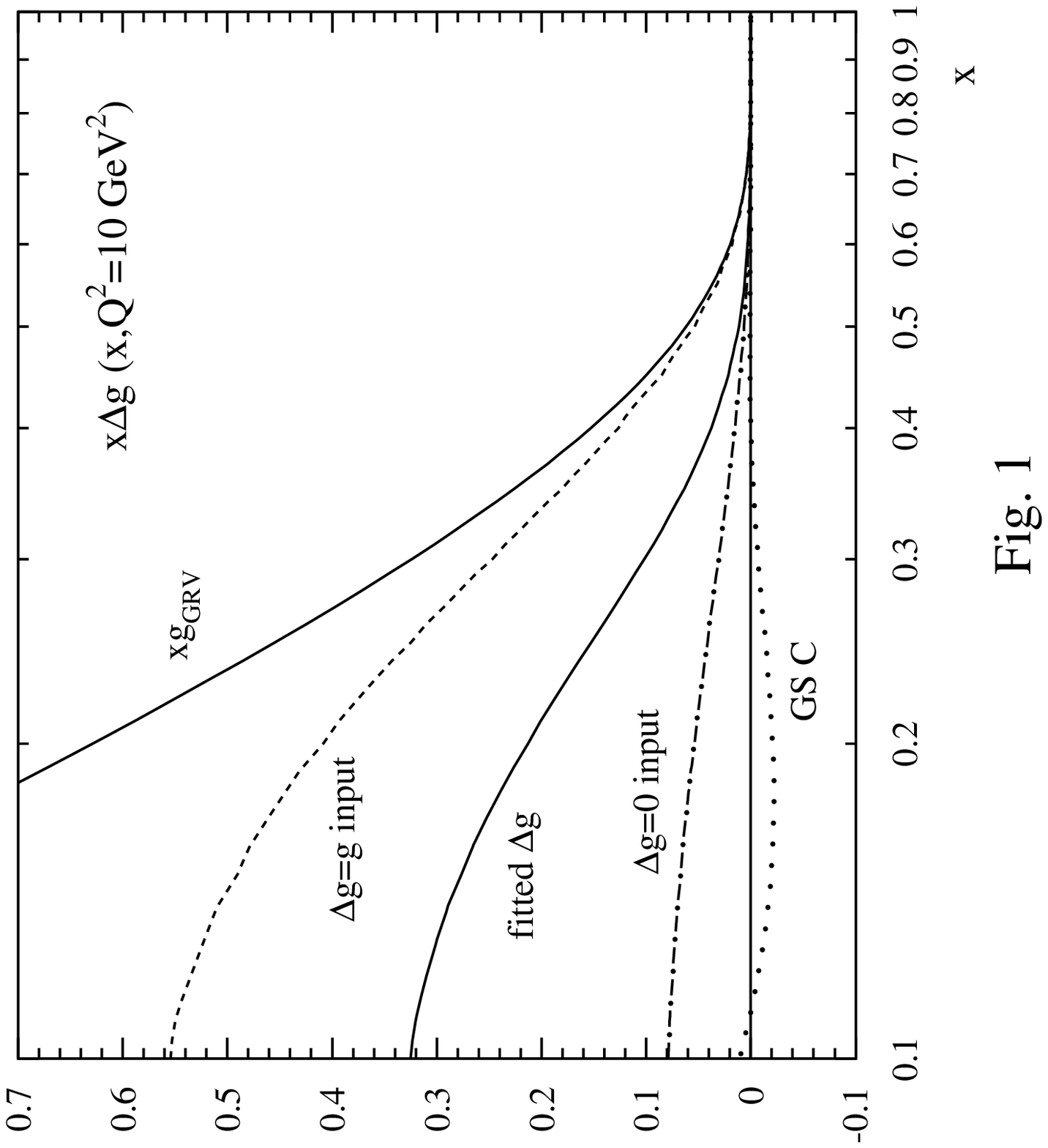,angle=90}
\newpage

\vspace*{-0.5cm}
\hspace*{-2.4cm}
\epsfig{file=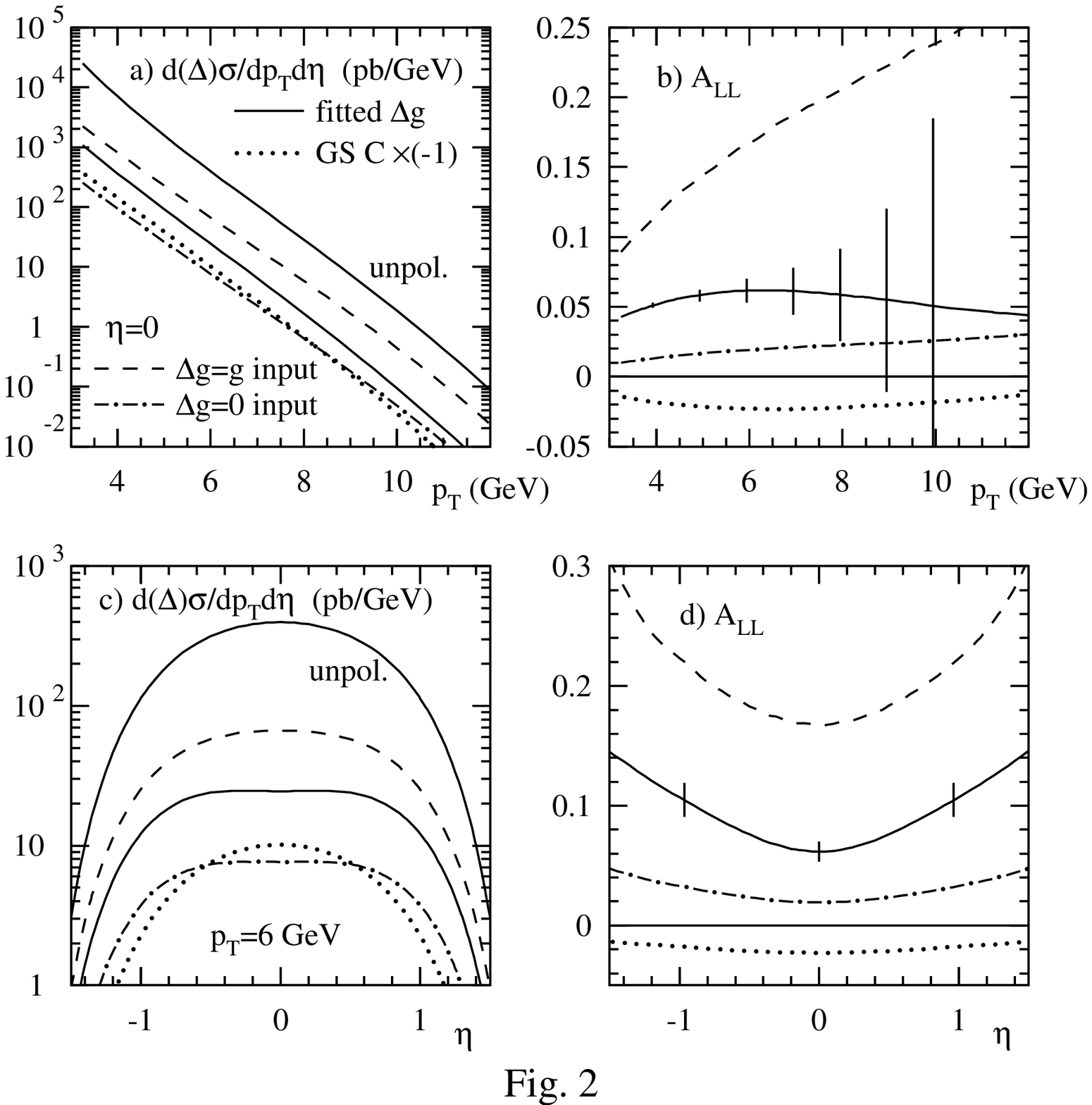,angle=0}
\newpage

\vspace*{-0.5cm}
\hspace*{-2.4cm}
\epsfig{file=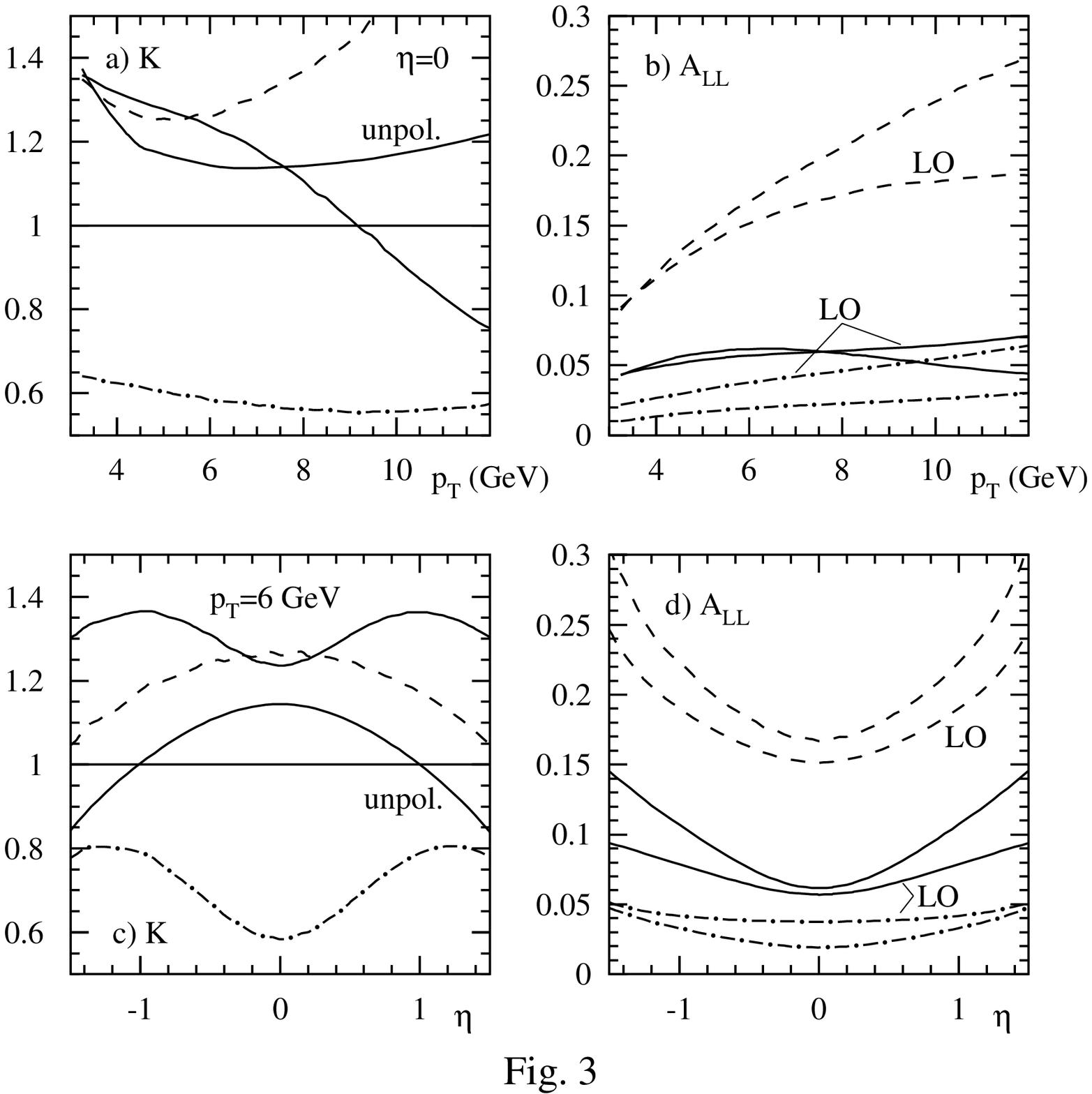,angle=0}
\newpage

\vspace*{-0.5cm}
\hspace*{-2.4cm}
\epsfig{file=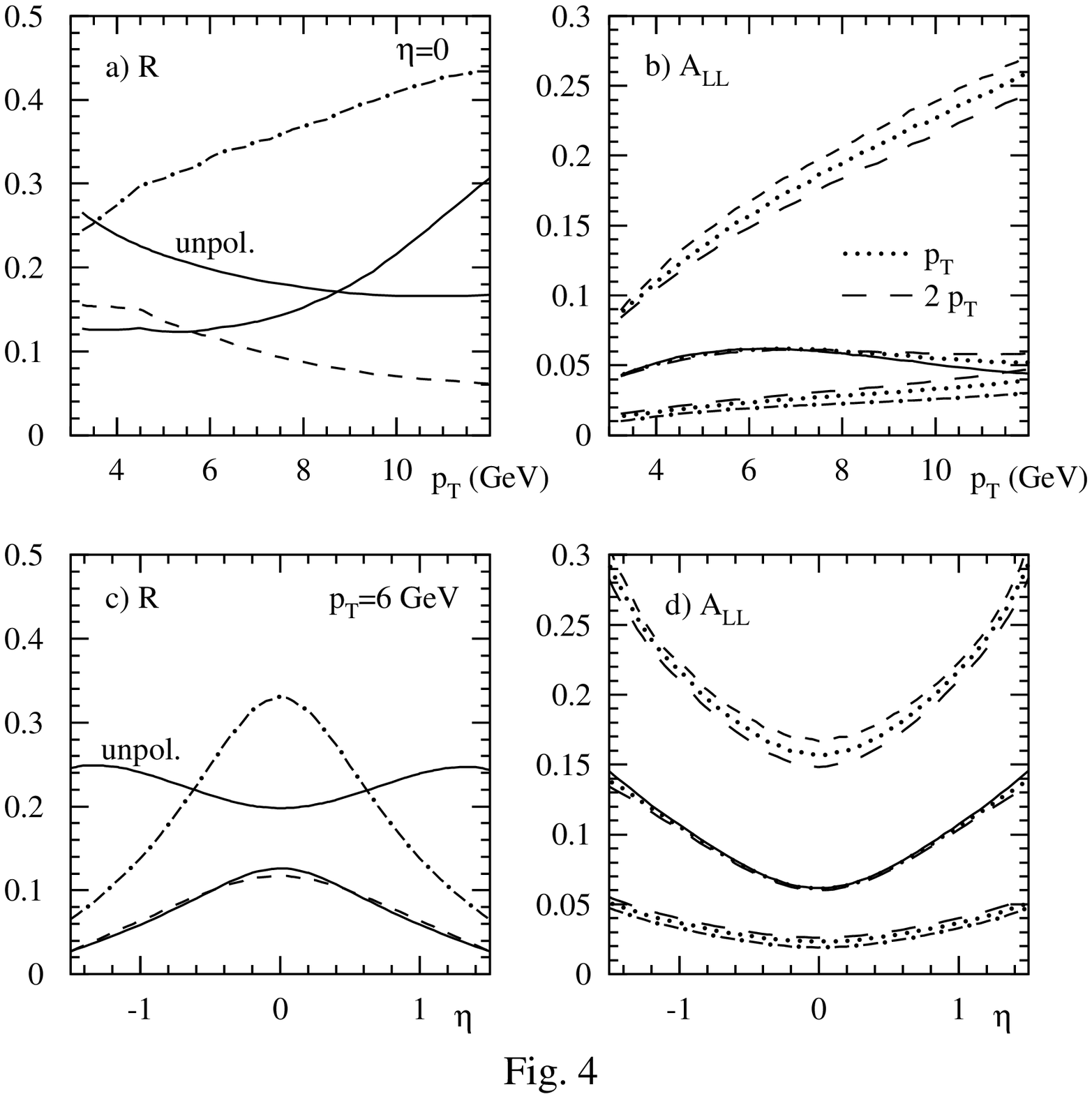,angle=0}

\end{document}